\documentstyle[eqsecnum,aps,twocolumn,epsfig,ilja1]{revtex}

\begin{document}
\draft
\title{Directional coupling of optical signals by odd dark beams with mixed
       phase dislocations}
\author{D. Neshev, A. Dreischuh, G.G. Paulus$^\ast$, H.
        Walther$^{\ast,\diamond}$}
\address{Sofia University, Dept. of Quantum Electronics,%\\
         5. J. Bourchier Blvd., BG-1164 Sofia, Bulgaria\\
         ( FAX: (+3592) 962 5276; e-mail: ald@phys.uni-sofia.bg )}
\address{$^\ast$Max-Planck-Institut f{\"u}r Quantenoptik, Institut f{\"u}r
         Laserphysik, %\\
         Hans-Korfermann-Str. 1, D-85748 Garching, Germany\\
         ( FAX: (+4989) 3290 5200; e-mail: ggp@mpq.mpg.de )}
\address{$^\diamond$Ludwig-Maximilians-Universit\"at, Sektion Physik,
         Am Coulombwall 1, D-85747 Garching, Germany\\
         ( FAX: (+4989) 2891 4142, E-mail: Prof.H.Walther@mpq.mpg.de )}

%\date{\today}
\maketitle

\begin{abstract}
Numerical simulations on the evolution of step-screw and edge-screw optical
phase dislocations in bulk saturable self-defocusing nonlinear media are
presented, with emphasis on their ability to induce steering waveguides for
signal beams (pulses). Two schemes for directional coupling of such signals,
both ensuring reasonable coupling-to-cross-talk efficiency ratio, are
investigated. The parameters useful for performance optimization of the
couplers are discussed.
\end{abstract}

%\pacs{{OCIS code:} 0190 4420; 0190 5940; 090 1760}

\section{Introduction}
Dark spatial solitons (DSSs) formed in bulk self-defocusing
Kerr-type nonlinear media (NLM) are able to guide co-propagating
probe beams/pulses \cite{ref1,Yuri,Law}. The physical mechanism
underlying is the intensity-dependent refractive-index change in a
plane perpendicular to the propagation direction. Weak signal
beams passing along these optically induced gradient waveguides
\cite{ref2} are subject to effective induced-phase modulation and
are trapped. In photorefractive NLM \cite{ref3} the situation is
even more spectacular \cite{ref4}. Because of the
wavelength-dependent material response, dark spatial solitons
generated at low powers but at a photosensitive wavelength are
able to guide and steer much more powerful `signal' beams at
non-photosensitive wavelengths. This is confirmed for photovoltaic
\cite{ref5}, biased \cite{ref6} and quasi-steady-state \cite{ref7}
photorefractive solitons and seems to hold also for incoherent
dark solitons \cite{ref8}.

Bright signal beams/pulses can be deflected by both electronically
\cite{ref9} and optically induced prisms \cite{ref10}. When they
are guided by DSSs, the same can be done by manipulating the
transverse dynamics of the dark beams. The transverse velocity of
an optical vortex soliton (OVS) has a radial and an angular
component arising from the transverse phase and intensity
gradients, respectively \cite{ref11,Rozas}. Two practical ways to
control the vortex rotation have their origin in the Guoy phase
shift on both sides of the background beam waist
\cite{Rozas,ref12} and in the interaction of ordered structures of
OVSs \cite{ref13} controlled by the topological charges. OVS
steering is demonstrated by superposition of a weak incoherent
background field \cite{ref14}. Operation of planar Y-junction
splitters for signal beams is demonstrated in both Kerr-type
\cite{ref15} and photorefractive NLM \cite{ref16} with pairs of
grey DSSs born from even initial conditions. The possibility to
branch a single input probe beam into ordered structures of
sub-beams by quasi-two-dimensional DSSs is demonstrated
numerically in Ref. \cite{ref17}. Other branching and steering
schemes can be realized by employing the inherent dynamics of ring
dark solitary waves \cite{ref18}, eventual NLM saturation
\cite{ref19} and/or anisotropy \cite{ref20}.

In this work we present numerical results on how one input optical
data channel can be linked to a selected output data channel by
means of an optically-induced steering waveguide. A high steering
speed of the induced waveguide can be obtained by using odd dark
beams (ODBs) of finite lengths \cite{ref23,ref24} with a suitable
choice of the mixed phase dislocation. Special attention is paid
to ensure high energy efficiency of the directional coupler and
short length of the interaction zone. Reconfiguration of the
coupler is proposed to be done by changing the type of the phase
dislocation reproduced by a multiple, active, single-voltage
controlled computer-generated hologram \cite{ref21}.

\section{Step-screw and edge-screw mixed phase dislocations}
The mixed phase dislocations considered consist of a
one-dimensional phase step of limited length, which ends, by
necessity, with pairs of phase semi-spirals with opposite
helicities. In this work two possible cases are considered. The
first phase dislocation of this type, which will be denoted
step-screw (SS), is described by the phase distribution
\begin{equation}
\label{gleich1}\Phi^{SS}_{\alpha,\beta}(x,y)=
\Delta \Phi\left\{\frac{1-\alpha}{2}\text{sgn}(y)
    -\frac{\beta}{\pi}\arctan\left(\frac{\alpha y}{x+b\beta}\right)\right\},
\end{equation}
where the parameters $\alpha$ and $\beta$ are defined as follows:
\begin{equation}
\label{gleich2}\alpha= \left\{
\begin{array}{ll}
0 & \text{for } |x|\leq b ,\\ 
1\text{ and }\beta=-1 & \text{for }x> b ,\\ 
1 \text{ and } \beta=1 & \text{for } x\le -b .
\end{array}
\right.
\end{equation}

\begin{figure}
\centerline{\epsfig{file=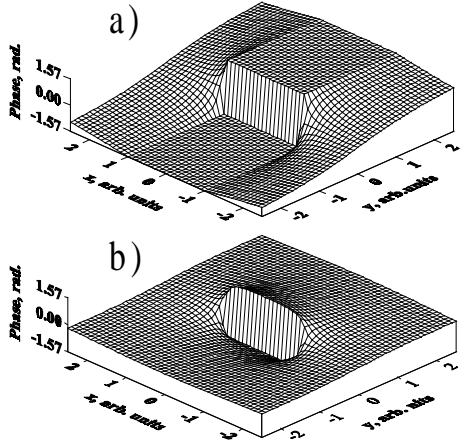,width=2.4in,clip=} }
\caption{(a) step-screw and (b) edge-screw mixed phase dislocations described
         by Eqs.~(2.1) and (2.3), respectively.}
\label{fig1}
\end{figure}

The phase distribution
\begin{equation}
\label{gleich3}\Phi^{ES}(x,y)=
\frac{\Delta \Phi}{2\pi}\left\{ \arctan\left( \frac{y}{x+b}\right)
-\arctan\left( \frac{y}{x-b} \right) \right\} ,
\end{equation}
will be referred to as edge-screw (ES). In
Eqs.~(\ref{gleich1})-(\ref{gleich3}) the quantity $\Delta\Phi$
stands for the magnitude of the step portion of the dislocation,
$2b$ for its length, and $x$ and $y$ denote the transverse
Cartesian coordinates parallel and perpendicular to the
dislocation. Surface plots of the SS and ES phase dislocations are
shown in Fig.~\ref{fig1}a and \ref{fig1}b, respectively.

The formation of mixed phase dislocations was first identified by
us in interferograms of decayed crossed one-dimensional DSSs in
the presence of moderate saturation of the nonlinearity
\cite{ref22}. From our present point of view the classification
\cite{ref23} of such formation generated via an instability as an
ES or SS phase dislocation cannot be definitive. Both distributions
shown in Fig.~\ref{fig1} are characterized by the same maximal
phase difference $\Delta\Phi$ and dislocation length $2b$, but
they differ in the rate $\partial\Phi/\partial\varphi |_{|x|>b}$
at which the phase changes azimuthally at their ends. In the
step-screw case this change rate is twice as high as in the
edge-screw case. In another work \cite{ref24} we found
experimentally and confirmed numerically that in the interval
$0.75\pi\leq \Delta\Phi\leq 1.25\pi$ the transverse steering
velocity of an ODB of SS type is inversely proportional to
$\Delta\Phi$. Because of the phase gradients across the edge
portion of the dislocation the ES one starts `immediately'
steering. The step portion of the SS dislocation, however, is
forced to steer by the outlying semi-helices. The process
stabilizes after a propagation distance of the order of $1L_{NL}$
(see Figs.~2 and 3 in Ref. \cite{ref23}), at a transverse velocity
2-3 times smaller than in the ES case. In view of this one more
general statement holds: The transverse velocity of an ODB with
mixed phase dislocation is inversely proportional to the absolute
value of the azimuthal phase-change rate
$\partial\Phi/\partial\varphi |_{|x|>b}$ outward its ends.
Therefore the identification of the type of ODBs generated as a
result of an instability is not very reliable.

\section{Initial conditions for the model}
The (2+1)-dimensional evolution of the ODBs in bulk homogeneous,
isotropic and saturable NLM is described by the nonlinear
Schr\"odinger equation
\begin{equation}
\label{Schroedinger}i\frac{\partial E_D}{\partial\zeta}+\frac{1}{2}
\Delta_{\perp} E_D - \frac{L_{Diff}}{L_{NL}}
{|E_D|^2E_D\over (1+s|E_D|^2)^\gamma}=0 ,
\end{equation}
where
$\Delta_\perp=\partial^2/{\partial\xi^2}+\partial^2/{\partial\eta^2}$.
The transverse spatial coordinates are normalized to the initial
dark beam width ($\xi=x/a$, $\eta=y/a$), and the nonlinear
propagation path length $\zeta$ is expressed in units of Rayleigh
diffraction lengths $L_{Diff}=ka^2$. Furthermore,
$L_{NL}=(kn_2I_0)^{-1}$ is the nonlinear length, $k$ is the wave
number inside the NLM, and the background beam intensity at its 
entrance isnormalized to that needed to form an 1D DSS of width $a$
($I_0=I_{DSS}^{1D}$). In order to maintain the
correspondence to the conditions in our previous experimental work
\cite{ref24,ref22}, where NLM with saturation was used, we adopted
the refractive-index correction
\begin{equation}
\label{gleich4} \Delta n=n_2|E_D|^2/(1+s|E_D|^2)^\gamma,
\end{equation}
with $s=0.3$ and $\gamma=3$. In a weak-signal approximation we
took into account the bright-signal beam diffraction and the
refractive-index changes it sees as induced by the ODB. In the NLM
its evolution is described by
\begin{equation}
\label{Schroedinger2}i\frac{\partial E_B}{\partial\zeta}+\frac{1}{2}
\frac{\lambda_B}{\lambda_D}
\Delta_\perp E_B- \sigma\frac{L_{Diff}}{L_{NL}}
{|E_D|^2E_B\over (1+s|E_D|^2)^\gamma}=0 .
\end{equation}

The coupling coefficient $\sigma$ depends on the nature of the
physical process of the optical nonlinearity (e.g. molecular
orientation, electronic response of bound electrons, etc.) and on
the experimental conditions (e.g. the polarizations of the waves).
The numerical simulations presented in this work refer to the
wavelength ratio $\lambda_B/\lambda_D\simeq 1$  and $\sigma=2$.
The slowly-varying electric-field amplitudes of the ODBs were
assumed to be $tanh$-shaped and of the form
\begin{equation}
\label{gleich5}E_D(x,y,z=0)=\sqrt{I_0}B(r_{1,0}(x,y))
\tanh\left[ \frac{r_{\alpha,\beta}(x,y)}{a}\right] e^{i\Phi(x,y)} ,
\end{equation}
where $\Phi(x,y)=\Phi^{SS}_{\alpha , \beta}(x,y)$ for an SS,
$\Phi(x,y)=\Phi^{ES}(x,y)$ for an ES phase dislocation, and
$r_{\alpha,\beta}(x,y)=\sqrt{\alpha(x+\beta b)^2+y^2}$ is the
effective radial coordinate. In order to avoid any influence of
the finite background beam of super-Gaussian form
$B(r_{1,0})=\exp\{-(\sqrt{{x^2+y^2}/{w^2}})^{14}\}$, its width was
chosen to exceed the maximal ODB deflection at $\zeta=10$ more
than 10 times. The signal beam is assumed to be $sech$-shaped and
equal in width to the ODB width $a$. Its intensity is chosen
hundred times weaker than $I_{DSS}^{1D}$, thus not disturbing the
dark beams by cross-phase modulation. Equations
(\ref{Schroedinger}) and (\ref{Schroedinger2}) are solved
numerically by modification of the beam propagation method over a
$1024\times1024$ grid.

\begin{figure}
\centerline{\epsfig{file=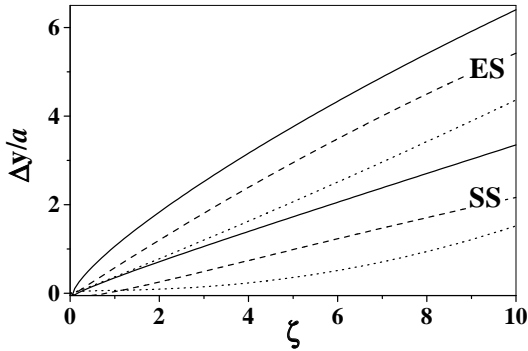,width=2.8in,clip=} }
\caption{Relative deflection of ODBs with edge-screw (ES) and
step-screw (SS) phase dislocations vs. nonlinear propagation path
length $\zeta =z/L_{NL}$ for $b/a=1.0$ (solid), $2.0$ (dashed), 
and $3.0$ (dotted) at $\Delta\Phi=\pi$.} 
\label{fig2}
\end{figure}

Figure~\ref{fig2} shows the deflection of ODBs with edge-screw
(ES) and step-screw (SS) phase dislocations vs. the nonlinear
propagation path length $\zeta=z/L_{NL}$ for $b/a=1.0$ (solid),
$2.0$ (dashed), and $3.0$ (dotted), normalized to the initial ODB
width $a$. The difference between the transverse dynamics of the
dark beams of different types is amazing. As already mentioned, it
is due to the presence (in the ES case) or absence (in the SS
case) of phase gradients across the one-dimensional portion of the
dislocation at the entrance of the NLM. Besides by the type of the
phase singularity, the transverse dynamics can be controlled in
three possible ways \cite{ref24}: by chaging $b/a$ ratio, the phase
step $\Delta\Phi$ and the background beam intensity. As it is seen 
in Fig.~\ref{fig2}, changing the ODB length-to-width ratio gives an
effective way to manipulate the ODB velocity. It is higher for the
short and slower for the long dislocations. The long ones however 
could experience a transverse instability. More strongly
pronounced is the increase of the velocity with decreasing the
magnitude of the step portion of the phase dislocation $\Delta \Phi$ 
\cite{ref24}, but we refrained from exploiting this. The reason is 
that during the steering process both types of ODBs inevitably 
become grey (and asymptotically disappear), which tends to destroy 
the coupler modeled. The numerical calculations performed showed 
similar but slightly weaker coupling efficiencies when the transverse
deflection is controlled by $\Delta\Phi$ than by $b/a$. All data
in this work refer to $\Delta\Phi(\zeta =0)=\pi$. Characteristic
of the evolution of the ES phase dislocation is that it disturbs the
background beam more weakly than the SS dislocation. In order to
emphasize this, the number of gray-scale levels in the frames
shown in Fig~\ref{fig3} are intentionally reduced in the same way.
The data refer to $\zeta=z/L_{NL}=4$ and $b/a=1$. The background
beam intensity in this simulation is chosen to correspond to that
of a 1D DSS ($I_{DSS}^{1D}$). The dashed line is intended to
underline that the ES ODB is stronger deflected than the SS ODB.

The background beam intensity could be additionally used to
controll the dynamics of the ODBs, even their deflection does not
strongly depend on it \cite{ref24}. The background beam intensity,
however, influences substantially the transverse beam profile and
the guiding properties of the induced waveguide. Better spatial
confinement could be expected at higher intensities. Numerical
simulations were made to confirm this statement. Depending on the
type of the dislocation (ES or SS) the highest encoupling
efficiencies were achieved for levels $(1.25\div 1.75)I_{DSS}^{1D}$.
Higher intensities tend to destroy the induced waveguide due to
the strong deformation of the dark beam profile under saturation
of the nonlinearity.

The main task of the numerical simulations presented bellow is to
find the proper parameter range for $b/a$ and beam intensity
in order to reach maximal coupling efficiency in the analyzed schemes 
for directional coupling.

\section{One-directional coupler}
\begin{figure}
\centerline{\epsfig{file=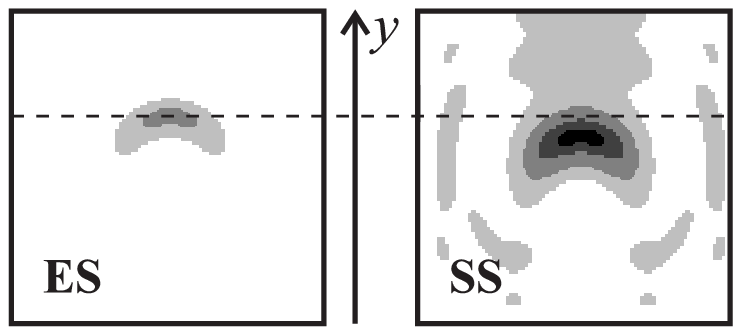,clip=} } 
\caption{Grey-scale plots of ES and SS ODBs with initial 
length-to-width ratio $b/a=1$ after nonlinear propagation path length 
$\zeta =4$. Some $3\%$ of the total computational area is shown. See 
text for details.}
\label{fig3}
\end{figure}

\begin{figure}
\centerline{\epsfig{file=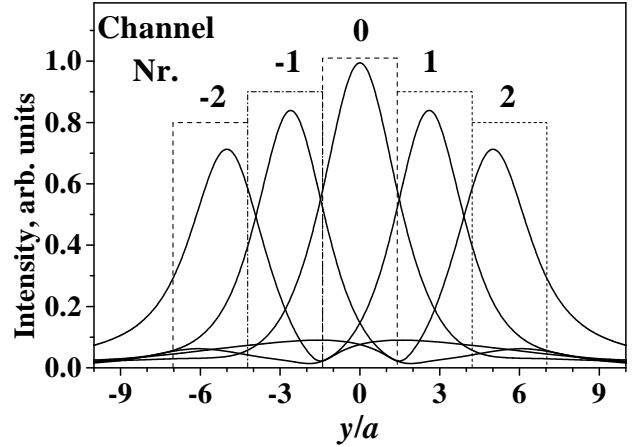,width=3.3in,clip=} }
\caption{Transverse profiles (along the steering direction $y$) of
the signal beams coupled to one of the five output channels
considered. The intensities are normalized to that of the signal
in the direct channel No.~$0$.} 
\label{fig4}
\end{figure}

As a first step in this work we modeled the coupling of an input
signal beam (pulse) to a desired output channel by optically
induced waveguides. Figure~\ref{fig4} shows the positions and the
forms of the weak signal beams guided by the corresponding dark
one after a nonlinear propagation path length $\zeta=10$. Channel
$0$ is assumed to be supported by an OVS \cite{Yuri,Law} with
backround intensity $1.5I_{DSS}^{1D}$, channels $1$ and $-1$ by
ODBs with SS phase dislocations of $b/a=1.5$ at $I=1.5I_{DSS}^{1D}$, 
whereas channels $2$ and $-2$ are considered to 
be induced by ODBs with ES phase dislocations with $b/a=2.5$ at
$I=1.25I_{DSS}^{1D}$. The change of the ODB steering direction in
the NLM (i.e. the switching between channels with positive and
negative numbers) is controlled by reversing the gradients of the
screw portions of the dislocations. First, real implementation of
this directional switching is feasible if there are reliable means
to change the phase distributions of the ODBs. The
electrically-controllable, multiply-active, computer-generated
holograms developed \cite{ref21} are well suited for doing that.
Because of the time needed to reconfigure the hologram, the
relatively rare redirection of the data streams addressed to a
desired channel will be preferable. Second, the energy efficiency
of the guiding and switching should be as high as possible at the
lowest possible cross-talk between the channels. The imaginary
information channels considered to be coupled to the exit of the
NLM are assumed to be rectangle-shaped with a width (height) 1.5
(3.0) times the input signal-beam FWHM. In Fig.~\ref{fig4} the
respective positions along the steering direction (the arrow in
Fig.~\ref{fig3}) are sketched as rectangles.

The results for the directional coupling and cross-talk
efficiencies obtained as ratios of the transmitted energy in a
particular channel to the total one are summarized in
Table~\ref{tbl1}. The ODB with an SS phase profile is assumed to
couple the input channel to channel No.~$1$, whereas this with an
ES dislocation redirects the input to channel $-2$. The maximal
efficiencies for channels $0$, $\pm 1$, and $\pm 2$ to be expected
are $74.7\%$, $70.7\%$, and $52.2\%$, respectively. In most
channels the cross-talk signal should remain well below $14\%$.
The ODB steering is accompanied by the creation of a leading dark
wave and a trailing bright peak on the background beam (see
Fig.~\ref{fig3} in Ref. \cite{ref23}) which are more strongly
pronounced in the SS case. The bright trailing peak pushes the
signal beam stronger. This qualitatively explains the relatively
high efficiency of the coupling to channels $+1$ and $-1$. Because
of the additional self-defocusing of this peak it appears more
effective to address channels $+2$ and $-2$ by ES ODBs. If the
out-lying channels are $2.25$ times larger than the input signal
beam width $a$, in the ES case the efficiency in coupling the 
signal to them can be increased by $3.6\%$ to $55.9\%$. Due to the
numerical discretization the overall accuracy in estimating the
energies is within $0.02\%$ and the coupling efficiencies in the
respective channels are correct to within $\pm 0.04\%$.

We should note here that because of the saturation of the
nonlinearity the width of the induced waveguide is larger than in
the case of pure Kerr nonlinearity. If a Kerr medium is considered
(i.e. $s=0$) the coupling efficiencies increase by $10\%$, whereas
the cross-talk coefficients become slightly weaker.

\section{Two-directional coupler}
\begin{figure}
\centerline{\epsfig{file=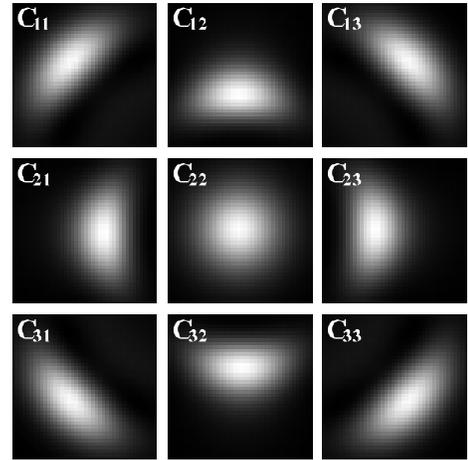,width=2.4in} }
\caption{Composition of grey-scale images of the signal beams at
the exit of the two-directional coupler. Each channel is addressed
separately and the coupler length is $\zeta =10$.} 
\label{fig5}
\end{figure}

For addressing a desired channel from an imaginary array of
information channels ($3\times 3$ in this work) we used the
squared-shaped channel configuration shown in Fig.~\ref{fig5}. The
data presented below refer to $\zeta =10$ and channel width and
height $1.5$ times the input signal beam FWHM. Best spatial
confinement of the signal beams is achieved at $1.5$ times higher
background beam intensity than that needed to form a 1D DSS of
width $a$. Besides being controlled by the type of the phase
dislocation (screw, ES, SS), the ODB transverse velocity and
deflection at the exit of the NLM are adjusted by the initial ODB
length-to-width ratio $2b/a$ ($b/a=1.5$ in the SS case; and 
$b/a=3.4$ in the ES case; both at $\Delta\Phi=\pi$). Let us denote 
the channels as $C_{IJ}$, where $I$ and $J$ are the respective row 
and column indices. The central output channel $C_{22}$, which is 
on axis with respect to the input one, is addressed by an OVS
\cite{Yuri,Law}, the side channels $C_{21}$, $C_{23}$, $C_{12}$,
and $C_{32}$ by ODBs with appropriately $\pm 90^\circ$ rotated SS
phase dislocations, whereas signal beam coupling to each of the
diagonal channels $C_{11}$, $C_{13}$, $C_{31}$, and $C_{33}$ is
performed by ODBs with ES phase dislocations rotated $\pm
45^\circ$. The calculated coupling efficiencies are summarized in
Table~\ref{tbl2}. The maximal efficiencies for the direct channel
$C_{22}$, the neighboring channel $C_{23}$, and the diagonal
channel $C_{11}$ to be expected are $61.6\%$, $49.6\%$, and
$47.1\%$ respectively, at cross-talk signals remaining well below
$12\%$. In Fig.~\ref{fig5} we composed grey-scale frames of the
signal beams as located on the imaginary array of information
channels considered. Coupling the input channel to information
channel $C_{11}$ (by an ES ODB) should result in undesired but
equal cross-signals in channels $C_{IJ}$ and $C_{JI}$ ($I\ne J$).
If the input signal is coupled to channel $C_{23}$ the same should
hold for channels $C_{1J}$ and $C_{3J}$. The data obtained
indicate that the discretization in our simulations leads to a
maximal inaccuracy in estimating the coupling efficiencies that is
within $\pm 0.04\%$. If pure Kerr nonlinearity is considered the
coupling efficiencies are $1\%$ to $8\%$ higher.

\section{Conclusion}
The numerical data on the nonlinear evolution of mixed step-screw
and edge-screw phase dislocations under equivalent conditions
shows an amazing difference between the transverse steering
dynamics of SS and ES ODBs. Such odd dark beams are able to induce
steering conduits for weak information beams (pulses) in bulk
nonlinear media. Due to this two different schemes for directional
coupling are proposed. The relatively high coupling and moderate
cross-talk efficiencies in both of them provide a reasonable basis
for further optimization and may open the way to constructing
parallel reconfigurable all-optical logical elements.

\acknowledgments D.N. thanks Vrije Universiteit Amsterdam for the
opportunity to complete part of the computational work at the
facilities of the Laser Center, VU, Amsterdam. A.D. is grateful to
the Alexander von Humboldt Foundation (Germany) for the award of a
fellowship and the opportunity to work in the stimulating
atmosphere of the Max-Planck-Institut f\"ur Quantenoptik
(Garching, Germany).

\begin{table}
\caption{One-directional coupler: Efficiencies (in \%) in
addressing channels $0$, $1$, and $-2$ by an OVS, SS ODB, and ES
ODB, respectively, and cross-talk efficiencies.} 
\label{tbl1}
%\centerline{                                    			%
\begin{tabular}{|r||c|c|c|c|c|c|}                       		%
\hline                                      				%
Channel         &  -2   & -1    & 0     & 1     & 2 & Energy    \\	%
guiding by      &       &       &       &       &   & losses    \\	%
\hline                                      				%
\hline                                      				%
OVS             & 0.3   & 9.5  &{\bf 74.7}& 9.5 & 0.3   & $<9.5$    \\	%
SS ODB          & 0.5   & 0.2   & 9.6   &{\bf 70.7}&5.6 & $<13.4$   \\	%
ES ODB          &{\bf 52.2}&14.1& 1.4   & 1.6   & 0.8   & $<29.9$   \\	%
                & (55.9)&       &       &       &       &	\\	%
\hline                                      				%
\end{tabular}%}
\end{table}

\begin{table}
\caption{Two-directional coupler: Efficiencies (in \%) in
addressing channels $C_{22}$, $C_{23}$, and $C_{11}$ by an OVS, SS
ODB, and ES ODB, respectively, and cross-talk efficiencies.}
\label{tbl2}
\begin{tabular}{|r||c|c|c|c|c|c|c|c|c|c|}
\hline
Channel &$C_{11}$ &$C_{12}$ &$C_{13}$& $C_{21}$ &$C_{22}$ &$C_{23}$
        &$C_{31}$ &$C_{32}$ &$C_{33}$ & Energy \\
guiding by&     &     &     &     &     &     &     &     &     & losses    \\ 
\hline 
\hline 
OVS    & 1.3 & 7.3 & 1.3  & 7.3 &{\bf 61.6}&7.3&1.3 & 7.3 & 1.3 &$<$4.0     \\ 
SS ODB & 0.2 & 2.4 & 12.6 & 0.1 & 6.1  &{\bf 49.6}&0.2&2.4& 12.6&$<$13.7    \\ 
ES ODB &{\bf 47.1}&5.5&0.7& 5.5 & 0.3  & 0.5  & 0.7 & 0.5 & 0.5 &$<$38.6    \\
\hline
\end{tabular}
\end{table}

\end{document}